\documentclass[prb,amsmath,superscriptaddress,twocolumn,amssymb,showpacs]{revtex4}
\usepackage{graphicx}
\usepackage{bm}
\usepackage{color}
\usepackage{dsfont}
\begin{document}
\title{Topological origin of edge states in two-dimensional inversion-symmetric insulators and semimetals}
\author{Guido van Miert}\affiliation{Institute for Theoretical Physics, Centre for Extreme Matter and Emergent Phenomena, Utrecht University, Princetonplein 5, 3584 CC Utrecht, The Netherlands}
\author{Carmine Ortix}\affiliation{Institute for Theoretical Solid State Physics, IFW Dresden, PF 270116, 01171 Dresden, Germany}
\affiliation{Institute for Theoretical Physics, Centre for Extreme Matter and Emergent Phenomena, Utrecht University, Princetonplein 5, 3584 CC Utrecht, The Netherlands}
\author{Cristiane Morais Smith}
\affiliation{Institute for Theoretical Physics, Centre for Extreme Matter and Emergent Phenomena, Utrecht University, Princetonplein 5, 3584 CC Utrecht, The Netherlands}
\affiliation{Wilczek Quantum Center, Zhejiang University of Technology, Hangzhou 310023, China}
\begin{abstract}
Symmetries play an essential role in identifying and characterizing topological states of matter. Here, we classify topologically two-dimensional (2D) insulators and semimetals with vanishing spin-orbit coupling using time-reversal ($\mathcal{T}$) and inversion ($\mathcal{I}$) symmetry. This allows us to link the presence of edge states in $\mathcal{I}$ and $\mathcal{T}$ symmetric 2D insulators, which are topologically trivial according to the Altland-Zirnbauer table, to a $\mathbb{Z}_2$ topological invariant. This invariant is directly related to the quantization of the Zak phase. It also predicts the generic presence of edge states in Dirac semimetals, in the absence of chiral symmetry. We then apply our findings to bilayer black phosphorus and show the occurrence of a gate-induced topological phase transition, where the $\mathbb{Z}_2$ invariant changes. 
\end{abstract} 
\pacs{03.65.Vf,73.20.-r, 73.22.-f}
\maketitle
\section{Introduction}The characterization of topological states of matter is a central topic in condensed-matter physics.\cite{HasanKane,QiZhang} A beautiful example is given by the Altland-Zirnbauer (AZ) table\cite{AZ}, which classifies topological insulators and superconductors depending on their dimensions and discrete symmetries.\cite{Ryu,Kitaev,Schnyder2} A 2D Chern insulator, for example, is characterized by a $\mathbb{Z}$ invariant and relies on the absence of time-reversal ($\mathcal{T}$) symmetry.\cite{TKNN} The presence of this symmetry is instead crucial for $\mathbb{Z}_2$ topological insulators that exhibit the quantum spin Hall effect.\cite{KaneMele} In these topologically non-trivial insulators, there is a one-to-one correspondence between the topological invariant and the number of gapless modes localized at the edge, known as the bulk-boundary correspondence.\cite{Laughlin} In Chern insulators the edge states are chiral, which means that they all propagate in the same direction, whereas topological $\mathcal{T}$ invariant insulators exhibit helical edge states, with electrons of opposite spins counterpropagating at the sample boundaries. 

There are, however, insulators and semimetals, which are topologically trivial according to the AZ-table, although they generally do exhibit edge states. A natural question is then whether these ``{\it trivial} '' edge states are related to a topological invariant that exists in the presence of a discrete or continuous symmetry. For 2D chiral systems, for instance, the existence of zero-energy edge states can be inferred from a 1D winding number.\cite{Hatsugai,Delplace} This explains the origin of trivial edge states in a minimal tight-binding model for graphene. More recently, this symmetry has been used to predict edge states in single-layer black phosphorus (sometimes also called phosphorene).\cite{Ezawa} However, this explanation is quite unsatisfactory for insulators and semimetals because  we do not expect the presence of a chiral symmetry: both in graphene and black phosphorus, the next-nearest neighbor hopping breaks the chiral symmetry. 

Here, we reveal the importance of inversion ($\mathcal{I}$)-symmetry, which has been overlooked in previous studies. We demonstrate that the existence of edge states in 2D crystalline insulators without spin-orbit coupling (SOC) in the presence of $\mathcal{T}$ and $\mathcal{I}$ symmetry is related to a one-dimensional (1D) $\mathbb{Z}_2$ invariant. We then apply this result to a generic toy model, and elucidate the relation between the edge states and this topological invariant. Moreover, we discuss the quantization of the edge charge in the absence of edge states. Finally, we apply our results to single and bilayer black phosphorus.
 
\begin{figure}[b]
\centering
\includegraphics[width=.45\textwidth]{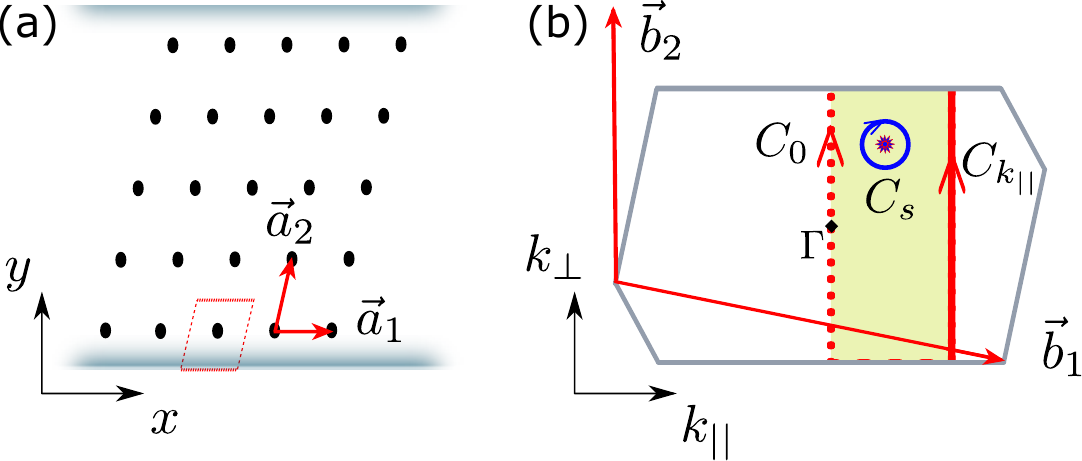}
\caption{\label{fig:lattices} (Color online) (a) Oblique lattice with lattice vectors $\vec{a}_1$ and $\vec{a}_2$. The red dashed rhomboid shows the unit-cell. (b) The corresponding 2D Brillouin zone, with the reciprocal lattice vectors $\vec{b}_1$ and $\vec{b}_2$. The red dashed line corresponds to the contour $C_0$, whereas the solid line corresponds to $C_{k_{||}}$, the blue circle denotes the contour $C_s$, and the enclosed area $S$ is shown in yellow.}
\end{figure}
\section{Symmetry of the Bloch Hamiltonian}
Let us consider a 2D crystalline insulator, described by a Bloch Hamiltonian $H(\vec{k})$. We assume negligible SOC, such that we can model the system using spin-less fermions, for which $\mathcal{T}^2=+1$. In the presence  of $\mathcal{I}$ and $\mathcal{T}$ symmetry, the Hamiltonian satisfies the following constraints:
\begin{align}
H(\vec{k})=\hat{I}H(-\vec{k})\hat{I},\quad H(\vec{k})=H(-\vec{k})^*,
\end{align}
where $\hat{I}$ is the inversion operator and $*$ denotes complex conjugation. Since we are interested in the existence of edge states, we put the system on a cylinder, as in Fig.~1(a), and limit ourselves to crystalline edges,  characterized by one of the lattice vectors $\vec{a}_1=(a_1,0)$. We will assign a topological invariant to the projected Bloch Hamiltonian $H_{k_{||}}(k_\perp)=H(k_{||},k_\perp)$,  where $k_{||}$($k_\perp$) denotes the momentum along (perpendicular to) the edge. To discuss the consequences of the constraints in Eq.~(1), we first briefly revisit the topology of band insulators in 1D.
\begin{table}[b!]
\begin{tabular}{|c|c|c||c|}
\hline
~~~~$\mathcal{I}$~~~~&~~~~$\mathcal{T}$~~~~&~~~~$\mathcal{I}\mathcal{T}$~~~~&~~~~1D~~~~\\
\hline
\hline
$\checkmark$&$\checkmark$&$\checkmark$&$\mathbb{Z}_2$\\
\hline
$\checkmark$&$\times$&$\times$&$\mathbb{Z}_2$\\
\hline
$\times$&$\checkmark$&$\times$&$0$\\
\hline
$\times$&$\times$&$\checkmark$&$\mathbb{Z}_2$\\
\hline
$\times$&$\times$&$\times$&$0$\\
\hline
\end{tabular}
\caption{Classification of 1D insulators in the presence or absence of $\mathcal{I}$, $\mathcal{T}$, and $\mathcal{T}\mathcal{I}$ symmetry, based on the Zak phase. Here, we consider spinless fermions, meaning $\mathcal{T}^2=+1$.}
\end{table}
\begin{table*}[t!]
\begin{tabular}{|l|l|l|l|}
\hline
(a)\quad $\xi(0)\xi(\vec{b}_2/2)=+1$&(b)\quad $\xi(0)\xi(\vec{b}_2/2)=+1$&(c)\quad $\xi(0)\xi(\vec{b}_2/2)=-1$&(d)\quad $\xi(0)\xi(\vec{b}_2/2)=-1$\\
\hline
\includegraphics[width=.24\textwidth]{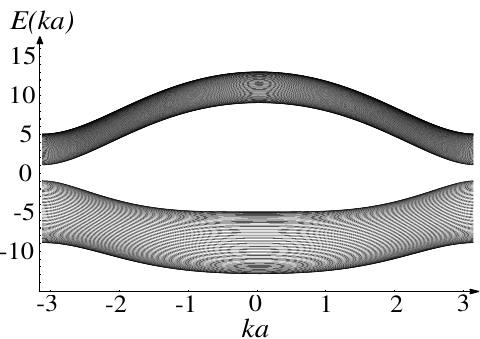}&\includegraphics[width=.24\textwidth]{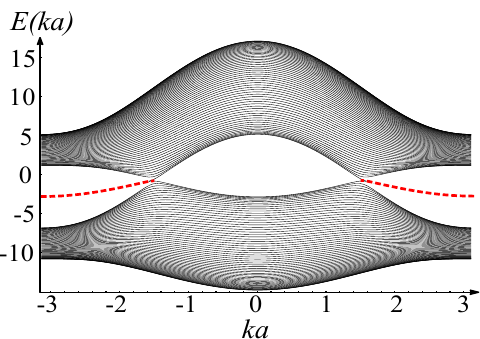}&\includegraphics[width=.24\textwidth]{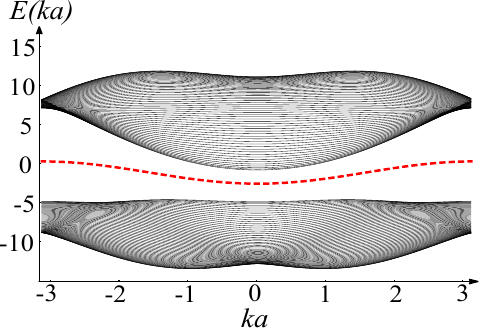}&\includegraphics[width=.24\textwidth]{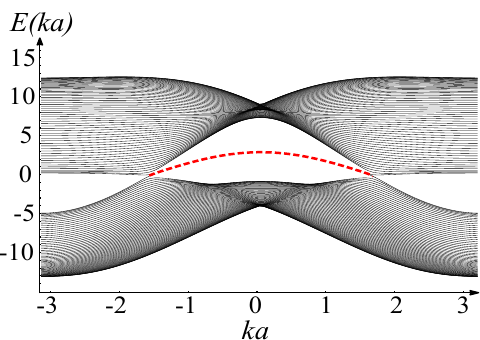}\\
\hline
\end{tabular}
\caption{(Color online) Four different band structures for a two-band model, with both $\mathcal{I}$ and $\mathcal{T}$ symmetry. Bulk (edge) states are indicated by solid (dashed) black (red) lines. (a) Spectrum of an insulator with a trivial $\mathbb{Z}_2$ invariant; (b) Dirac semimetal with a trivial $\mathbb{Z}_2$ invariant; (c) Again an insulator, but now with a non-trivial $\mathbb{Z}_2$ invariant;  (d) Dirac semimetal with a non-trivial $\mathbb{Z}_2$ invariant. The parameters used for these four dispersion relations are given in App.~C.}
\end{table*}

\section{Band topology in 1D} According to the AZ-table, insulators in 1D, belonging to classes A, AI, and AII, are trivial. Despite this fact, band topology does play an important role for these crystalline insulators. In particular,  the surface charge at the end of such an insulator is directly related to the Zak phase\cite{Zak,Vanderbilt} 
\begin{align}
 \gamma&=\int_{-\pi/a}^{\pi/a}\mathrm{d}k A(k),
\end{align}
where $a$ is the lattice constant and $A(k)=\sum_{i\in \textrm{occ}}\langle u_{k,i}|i\nabla_k|u_{k,i}\rangle$ denotes the Berry connection, with $|u_{k,i}\rangle$ the periodic part of the Bloch wave function with band index $i$, momentum $k$, and ``occ'' denotes the set of occupied bands. Generically, the Zak phase is not quantized. However, in the presence of 1D $\mathcal{I}$-symmetry, one finds $\gamma=-\gamma\textrm{ mod }2\pi$ (see Appendix~A). Hence, either $\gamma=0$ or $\gamma=\pi$, and as such it defines a $\mathbb{Z}_2$ invariant.\cite{Bernevig} Moreover, we do not need to compute the integral in Eq.~(2); instead, we can write
\begin{align}
e^{i\gamma}=\prod_{i\in \textrm{occ}}\xi_i(0)\xi_i(\pi/a)\in\mathbb{Z}_2,
\end{align}
where $\xi_i(k_\textrm{inv})=\langle \Psi_{k_\textrm{inv},i}|\hat{I}|\Psi_{k_\textrm{inv},i}\rangle$ is the parity of the full Bloch wave function $|\Psi_{k_\textrm{inv},i}\rangle$ at the $\mathcal{I}$ invariant momenta $k_\textrm{inv}$. By considering $\mathcal{T}$-symmetry, one can construct five different classes. In 1D, three of these allow for a $\mathbb{Z}_2$ classification, based on the Zak phase, see Table~I and Appendix~A. Next, we employ this invariant to characterize a 2D system.

\section{$\mathbb{Z}_2$ classification} Let us now return to the set of 1D Hamiltonians $H_{k_{||}}(k_\perp)$ parameterized by $k_{||}$. Note that for $k_{||}=0$, the symmetry constraints given in Eq.~(1) are inherited. In particular, the 2D $\mathcal{I}$ symmetry yields $H_0(k_\perp)=\hat{I}H_0(-k_\perp)\hat{I}$. Therefore, the associated Zak phase $\gamma(0)$ is quantized, and defines a $\mathbb{Z}_2$ invariant\cite{ortix,hatsugai3},
\begin{align}
\chi_1=e^{i\gamma(0)}=\prod_{i\in \textrm{occ}}\xi_i(0)\xi_i(\vec{b}_2/2).
\end{align}
Here, $\vec{b}_2$ is the reciprocal-lattice vector pointing in the direction perpendicular to the edge, see Fig.~1(b), and the subscript in $\chi_1$ reminds us of the fact that this invariant is associated with an edge parallel to $\vec{a}_1$. If the edge would be along $\vec{a}_2$, then we should simply consider the invariant $\chi_2=\prod_{i\in \textrm{occ}}\xi_i(0)\xi_i(\vec{b}_1/2)$. The definition of the invariant in Eq.~(4) is, strictly speaking, only valid for $k_{||}=0$. However, we can express the difference in the Zak phases $\gamma(k_{||})-\gamma(0)$ as an integral of the Berry connection $\vec{A}$ along the contour $C_{k_{||}}-C_0$, see Fig.~1(b). Using Stokes' theorem, this  can be rewritten as a surface integral of the Berry curvature $F(k_{||},k_\perp)=\partial_{k_{||}}A_\perp-\partial_{k_\perp}A_{||}$, see Fig.~1(b). Since the two symmetry constraints in Eq.~(1) ensure that $F=0$ (see Appendix~B), we find $\gamma(k_{||})=\gamma(0)$. It follows that even in the absence of chiral symmetry, we can still associate a $\mathbb{Z}_2$ invariant with each of the 1D Hamiltonians $H_{k_{||}}$.

So far, we have implicitly assumed that $H_{k_{||}}$ and $H_0$ are adiabatically connected, such that $F$ is actually well defined within the yellow area in Fig.~1(b). Therefore, our proof does not apply to systems with band-crossing points, like semimetals. In the latter case, we can modify our proof, by assuming that the band-crossing point is located inside the contour $C_s$, as indicated in Fig.~1(b). Then, we can apply Stokes' theorem to the new contour $C_{k_{||}}-C_0-C_s$, which yields
\begin{align}
\gamma(k_{||})-\gamma(0)&=\oint_{C_s}\mathrm{d}\vec{k}\cdot\vec{A}(k_x,k_y)=j \pi,
\end{align}
where $j$ is an integer. This result follows from the quantization of the Zak phase $\gamma_s=\oint_{C_s}\mathrm{d}\vec{k}\cdot\vec{A}(k_x,k_y)$  associated with the band crossing in multiples of $\pi$ (see Appendix~A). Thus, for a semimetal,  the Zak phase $\gamma(k_{||})$ is quantized, but changes by $j\pi$ as one encloses a $j\pi$-Berry phase degeneracy. Henceforth, for both 2D insulators and semimetals, we can introduce a topological $\mathbb{Z}_2$ invariant $\chi_1$ protected by $\mathcal{I}$ and $\mathcal{T}$ symmetry, which reflects the fact that the Zak phase $\gamma(k_{||})$ is quantized for all values of $k_{||}$. 

\section{Bulk-Boundary Correspondence} Next, we discuss the bulk-boundary correspondence, which relates the $\mathbb{Z}_2$ invariant $\chi_1$ to the edge behavior. For this purpose, we consider a two-band toy model on an oblique lattice. Specifically, we study a system with edges as in Fig.~1(a), for which every site hosts an $s$ and a $p$ orbital. The corresponding bulk Hamiltonian can be written as $H(\vec{k})=h_I(\vec{k})\mathds{1}+h_y(\vec{k})\sigma_y+h_z(\vec{k})\sigma_z$,
where $h_I$, $h_y$, and $h_{z}$ are real-valued even functions due to the constraints in Eq.~(1), and $\sigma_i$ are the Pauli matrices. Moreover, in this basis the inversion operator is given by $\hat{I}=\sigma_z$, such that $\hat{I}|s\rangle=|s\rangle$ and $\hat{I}|p\rangle=-|p\rangle$. At half-filling, we have $\xi(\vec{k}_{\textrm{inv}})=sgn[h_z(\vec{k}_{\textrm{inv}})]$. 
Hence, we can express the $\mathbb{Z}_2$ invariant as
\begin{align}
\chi_1&=sgn[h_z(0)]sgn[h_z(\vec{b}_{2}/2)].
\end{align}
If one only includes nearest-neighbor hopping, the bulk Hamiltonian is specified by
\begin{align*}
h_I+h_z&=e_s+2t_{s,1}\cos(\vec{k}\cdot \vec{a}_1) +2t_{s,2}\cos(\vec{k}\cdot \vec{a}_2),\\
h_I-h_z&=e_p+2t_{p,1}\cos(\vec{k}\cdot \vec{a}_1) +2t_{p,2}\cos(\vec{k}\cdot \vec{a}_2),\\
h_{y}&=2t_{sp,1}\sin(\vec{k}\cdot\vec{a}_1) +2t_{sp,2}\sin(\vec{k}\cdot\vec{a}_2),
\end{align*}
where $e_s$ and $e_p$ denote the on-site energies, $t_{s,i}$ and $t_{p,i}$ are the nearest-neighbor-hopping parameters in the direction $i$, and $t_{sp,i}$ is the hybridization among $s$ and $p$ orbitals. In Table~II, we display four  spectra, which are realized for values of the above parameters chosen {\it ad hoc}, in a way to provide an example of the four qualitatively different scenarios (see Appendix~C). We will shortly discuss below how their distinct features can be understood in terms of the $\mathbb{Z}_2$ invariant.\\
{\it (i) Gapped bulk with a trivial $\mathbb{Z}_2$ invariant:}\\
For $\chi_1=+1$, and in the absence of band-crossing points, the Zak phase $\gamma_{k_{||}}=0$ for all $k_{||}$. In this case, the trivial value of the Zak phase for all values of the momentum is reflected in the absence of edge states. This behavior is confirmed by plotting the spectrum for an insulator with $\chi_1=+1$ in Table~II(a), which only shows bulk states.\\
{\it (ii) Gapless bulk with a trivial $\mathbb{Z}_2$ invariant:}\\
The combination of a trivial $\mathbb{Z}_2$ invariant, $\chi_1=+1$, and two $\pi$-Berry phase band-crossing points yields a trivial Zak phase $\gamma(k_{||})=0$ for momenta adiabatically connected to $0$, and a non-trivial Zak phase $\gamma(k_{||})=\pi$ for $k_{||}$ outside this region. Hence, for momenta contained in the latter region, the bulk topology gives rise to edge states. This is indeed the case for the spectrum shown in Table~II(b), which corresponds to a Dirac semimetal with $\chi_1=+1$. \\
{\it (iii) Gapped bulk with a non-trivial $\mathbb{Z}_2$ invariant:}\\
For a non-trivial insulator, with $\chi_1=-1$, the Zak phase $\gamma(k_{||})=\pi$ for all $k_{||}$. The non-trivial Zak phase manifests itself via the presence of edge states for all momenta. This is verified in the example shown in Table~II(c) (where $\chi_1=-1$), which features in-gap edge states for all momenta.\\
{\it (iv) Gapless bulk with a non-trivial $\mathbb{Z}_2$ invariant:}\\
Finally, we consider a gapless system with a non-trivial invariant, $\chi_1=-1$. Then, the Zak phase $\gamma_{k_{||}}=\pi$ for the values of $k_{||}$ that are adiabatically connected to $k_{||}=0$, whereas for $k_{||}$ outside this region $\gamma(k_{||})=0$. Hence, we expect the presence of edge states for $k_{||}$ that are adiabatically connected to $k_{||}=0$. The band structure shown in Table~II(d) confirms this expectation. 

Thus, for insulators the $\mathbb{Z}_2$ invariant determines the presence or absence of edge states, whereas for a Dirac semimetal it encodes whether the two Dirac cones are connected by edge states which go through zero or $\pi$.  In particular, we might view Dirac semimetals as systems that interpolate between different $\mathbb{Z}_2$ insulators. Moreover, our results show that edge states are a robust feature of Dirac semi-metals. The presence of edge states is guaranteed, as long as the two Dirac cones are separated in the one-dimensional Brillouin zone.\\
\begin{figure}[t!]
\centering
\includegraphics[width=.48\textwidth]{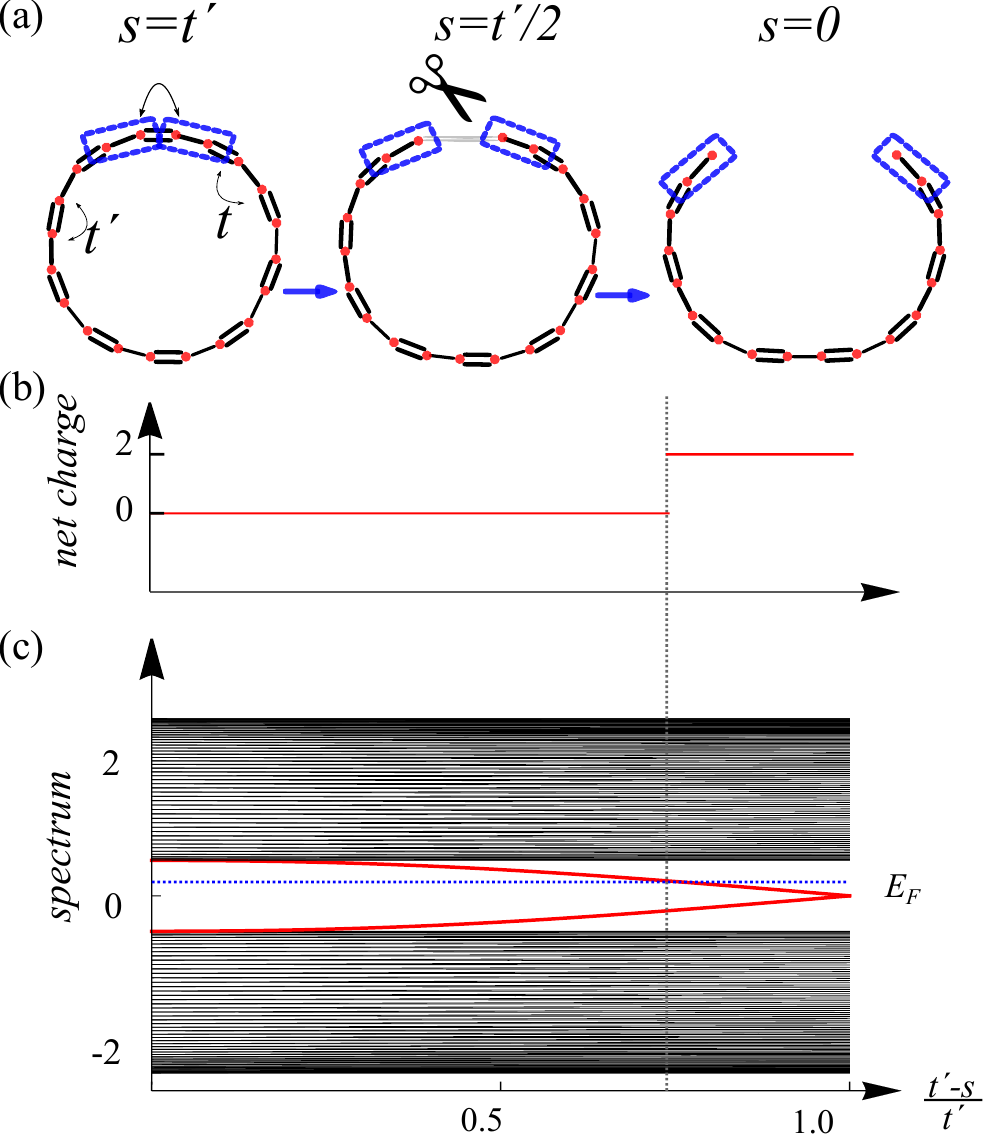}
\caption{\label{fig:lattices} (Color online) (a) Sketch of a periodic Su-Schrieffer-Heeger chain, which is adiabatically deformed into a chain with open ends. The parameter $s$ is the hopping parameter that connects the two blue unit-cells. (b) Net charge present on the chain as a function of the hopping parameter $s$, which is adiabatically turned off. (c) Corresponding spectrum. Here, we have considered a chain containing $100$ unit cells.}
\end{figure}
\section{Edge Charge Quantization}
We now explain the deep connection between the $\mathbb{Z}_2$ invariant $\chi_1$ and the emergence of in-gap states. It is a direct consequence of the boundary-charge theorem,\cite{Vanderbilt} which states that for a one-dimensional crystalline spin-degenerate insulator the surface charge $\sigma$ is well defined modulo $2e$, and is given by
\begin{align}
\sigma&=\pm\left(\frac{e\gamma}{\pi}+e\sum_{j=1}^\mathcal{N}Z_j u_j + e\sum_{j\in \textrm{surf}}Z_j\right).
\end{align}
Here, $e$ is the electron charge, $\gamma$ is the Zak phase, $\mathcal{N}$ is the total number of atoms within a unit cell, and  $Z_j$ and $u_j$ denote, respectively, the atomic number and position of the $j$th atom within the unit cell. The second term vanishes for $\mathcal{I}$-symmetric insulators. Moreover, the third term counts the total ionic charge of the atoms contained in the set ``surf''. These are the atoms that remain at the edges when tiling the finite system with unit cells. Note that this term precisely cancels the unit-cell ambiguity stemming from the first two terms. The $\pm$ refers to the left and right surface charge. In the presence of $\mathcal{I}$-symmetry, $\sigma$ is quantized to $0$ or $e$ modulo integer multiples of $2e$. In principle, this $2e$ ambiguity allows for different right- and left-surface charges; however, when the open chain is $\mathcal{I}$-symmetric, we find that $\sigma_\textrm{left}=\sigma_\textrm{right}$. Hence, the net charge present on the open chain is then $Q_\textrm{net}=\sigma_\textrm{right}+\sigma_\textrm{left}=2\sigma_\textrm{right}$ modulo $4e$. The relation between the net charge and in-gap states can be brought to light using the following adiabatic continuity argument, which we illustrate for a Su-Schrieffer-Heeger chain that we put in a ring geometry, as depicted in Fig.~2(a). Electrons on the chain can hop with hopping parameters $t$ and $t'$, corresponding to the single and double bonds. For simplicity, we set  $t=-0.8$eV and $t'=-1.2$eV. In addition, we consider a weak link with hopping parameter $s$, that can be varied. The presence of the weak link defines a preferential unit cell, which contains a single bond in our example. The Zak phase is then $\gamma=\pi$. From the boundary-charge theorem, it follows that, for an open chain ($s=0$), both ends of the chain will have a surface charge $\sigma=e$. Therefore, the net charge will be $2e$ modulo $4e$. This implies, that if one adiabatically changes the weak-link hopping parameter from $s=t'$ to $s=0$, an odd number of spin-degenerate states crosses the Fermi level: an in-gap state must have appeared, see Figs.~2(b) and (c). This does not guarantee that in-gap states will be present at the end of the adiabatic deformation ($s=0$). In-gap states can dissolve into the bulk, whereas they are pinned at zero energy in systems with particle-hole symmetry. 

\begin{figure}[b!]
\centering
\includegraphics[width=.48\textwidth]{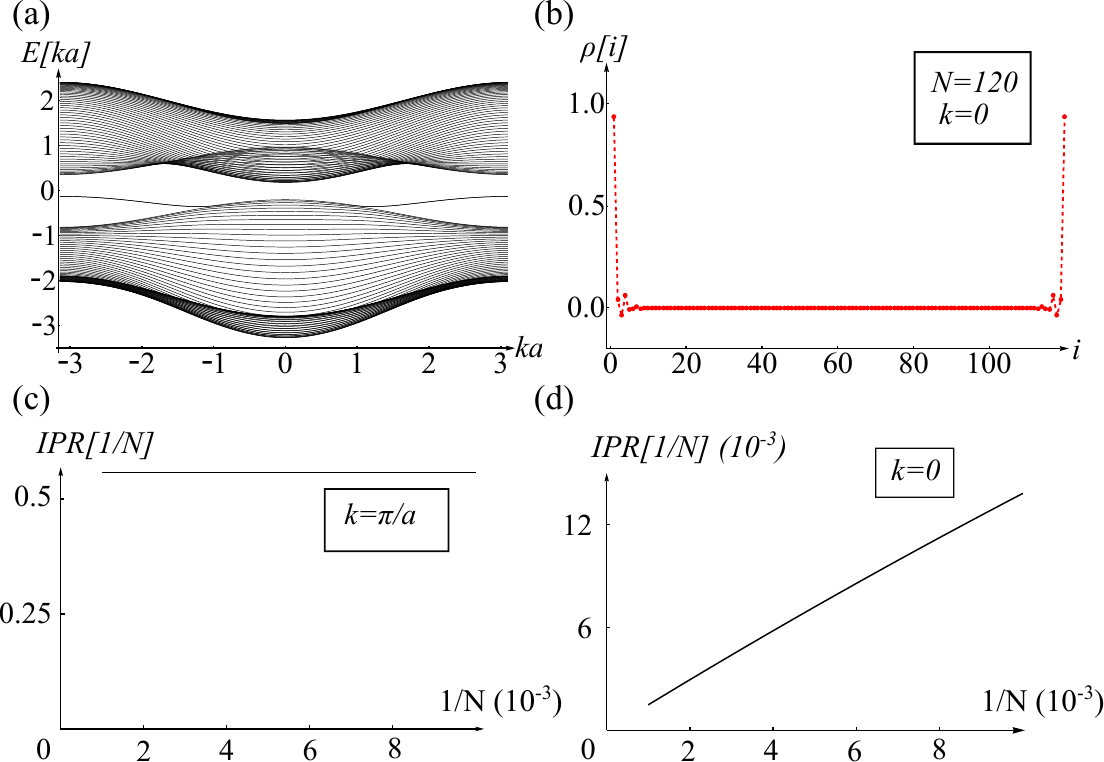}
\caption{\label{fig:lattices} (Color online) (a) Spectrum of a two-band model with $\chi_1=-1$. (b) Net charge density $\rho(i)$ as a function of the lattice site for $k=0$ and chain length $N=120$. (c) IPR of one of the in-gap states at $k=\pi/a$. (d) IPR of the highest-valence band states at $k=0$.}
\end{figure}

To demonstrate this statement, we consider the band structure shown in Fig.~3(a). The corresponding bulk Hamiltonian is still described by the two-band toy model, with $\chi_1=-1$ (see Appendix~D for the details). The in-gap edge states are absent for $k_{||}=0$, while for sufficiently large values of $k_{||}$, they emerge from the bulk. Hence, the in-gap state that must have traversed the band gap leaves no clear trace in the spectrum at $k_{||}=0$. Despite the trivial spectrum at $k_{||}=0$, we find that the non-trivial topology is captured by the net charge distribution $\rho(i)$, as shown in Fig.~3(b), thereby verifying the validity of the boundary-charge theorem, Eq.~(7). In particular, it follows that for 2D insulators in the presence of $\mathcal{I}$- and $\mathcal{T}$-symmetry, the edge charge associated with $H_{k_{||}}$ is either $0$ or $e$, as numerically confirmed in Fig.~3(b). 

To verify the absence of bound states at $k_{||}=0$, we have analyzed the inverse participation ratio (IPR), which quantifies over how many sites a particular state is distributed.\cite{Wegner} For a given state $|\Psi\rangle$ in a 1D system, the IPR is defined as
\begin{align}
IPR(|\Psi\rangle)=\sum_{i,\alpha}|\langle i,\alpha|\Psi\rangle|^4,
\end{align}
where $|i,\alpha\rangle$ denotes the state localized at site $i$, and $\alpha$ labels the orbital. For a proper bulk state, the IPR as a function of the chain length $N$ should be proportional to $1/N$, whereas for an edge state the IPR goes to a constant value. In Fig.~3(c), we plot the IPR for one of the in-gap states at $k_{||}=\pi/a$. We can clearly see that this is indeed an edge state. In contrast, in Fig.~3(d) we plot the IPR for the highest-valence band state at $k_{||}=0$, which can be identified as a bulk state. 
\begin{figure}[b!]
\centering
\includegraphics[width=.48\textwidth]{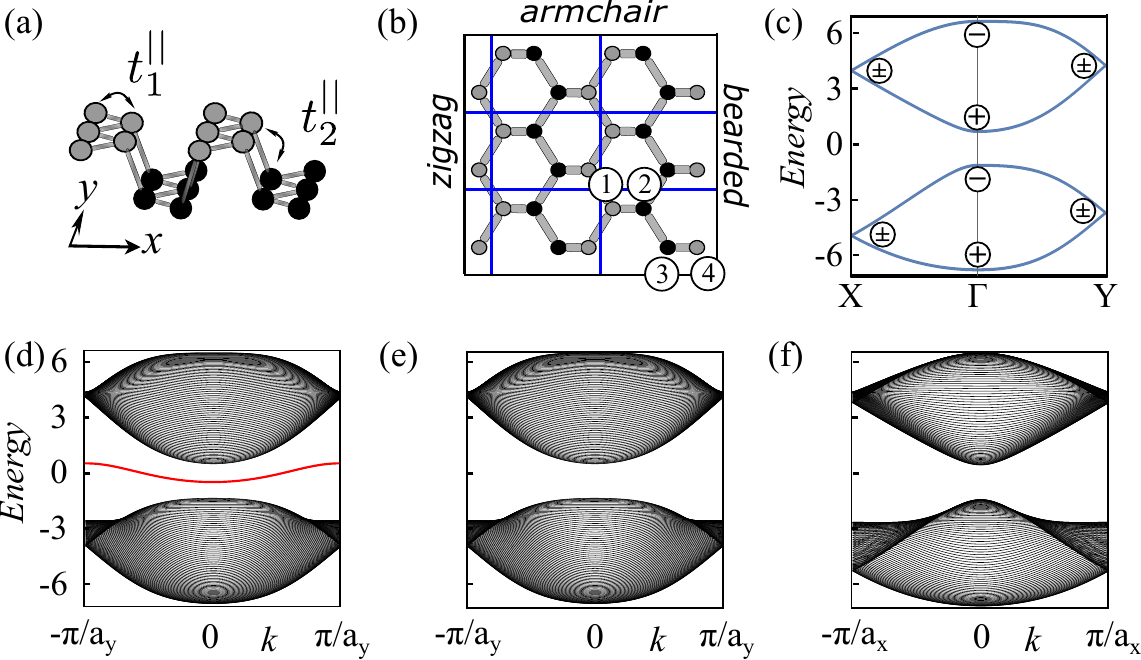}
\caption{\label{fig:lattices} (Color online) (a) Bird's eye view of single-layer black phosphorus. (b) Top view. (c) Bulk band structure along high-symmetry lines, where the $\pm$ signs indicate the parities of the Bloch waves.  (d), (e), and (f) Spectra of ribbons single-layer black phosphorus with zigzag, bearded, and armchair edges, respectively.}
\end{figure}
\begin{figure*}[t!]
\centering
\includegraphics[width=\textwidth]{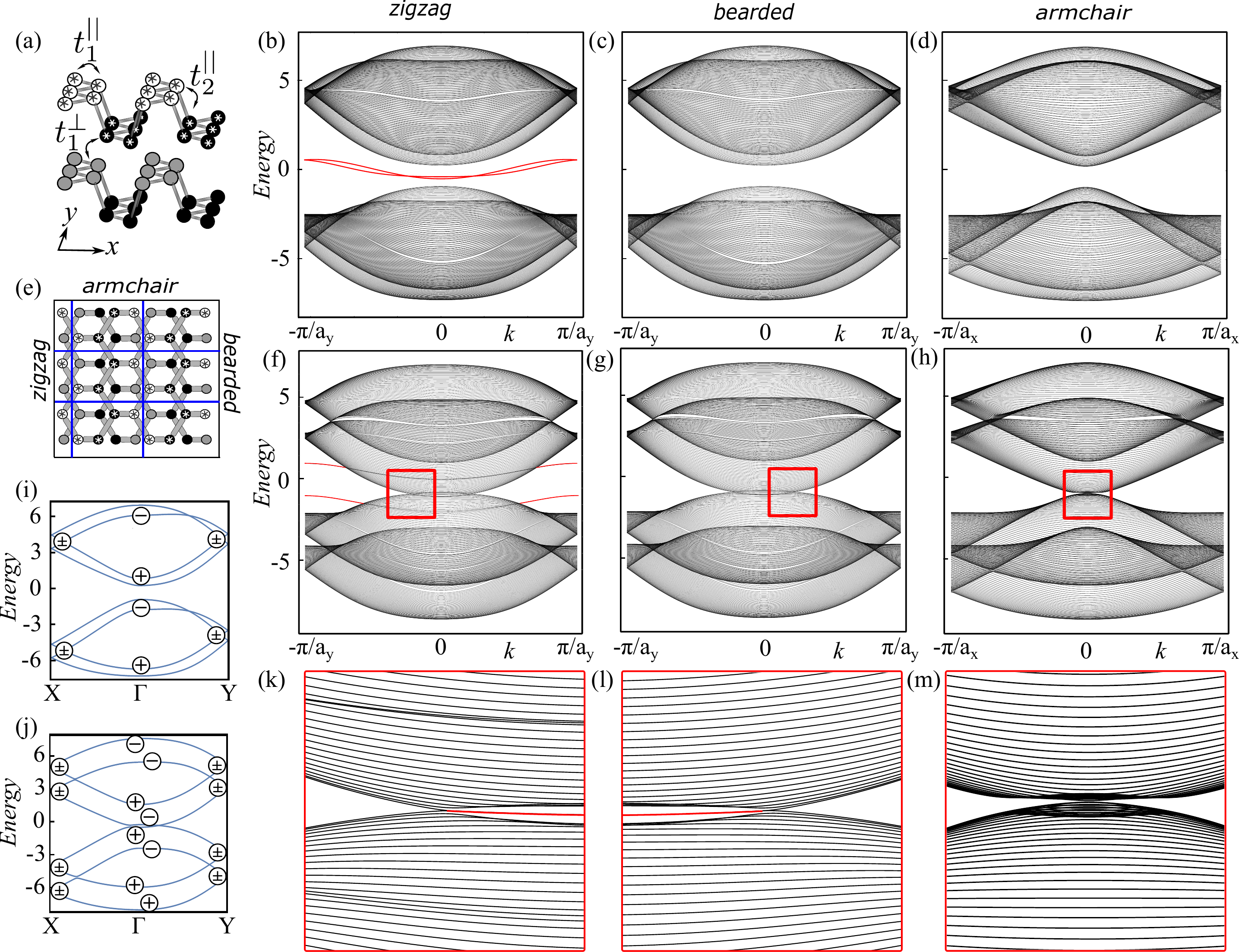}
\caption{\label{fig:lattices} (Color online) (a) Bird's eye view of bilayer black phosphorus. (b) , (c), and (d) Spectra of a ribbon of bilayer black phosphorus with zigzag, bearded and armchair edges, respectively. (e) Top view of bilayer black phosphorus. (f), (g), and (h) Spectra of a biased ribbon of bilayer black phosphorus ($\Delta V=2$eV), with  zigzag, bearded and armchair edges, respectively. (i) and (j) Bulk band structure along high-symmetry lines, for bilayer black phosphorus with $\Delta V=0$ and $\Delta V=2$eV, respectively. (k), (l), and (m) close-up of Figs. (f), (g), and (h) around the band-crossing point.}
\end{figure*}

\section{Few-layer Black Phosphorus} Next, we apply our results to an actual 2D material: black phosphorus, which consists of stacked sheets coupled by Van der Waals forces. Recently, it has been possible to isolate individual layers of black phosphorus (phosphorene) by mechanical exfoliation.\cite{Li,Buscema,Koenig,Liu,Xia} In Figs.~4(a) and (b), we display a sketch of such a layer. Like graphene, phosphorene has a honeycomb lattice; however, the bonds in phosphorene result from $sp^3$ hybridization, which leads to the puckered structure.  Phosphorene is a semiconductor, with a band gap that goes from $0.3$eV to $2$eV, depending on the number of layers. This sizable gap makes it a very promising material for electronic applications.

Here, we demonstrate that this novel material features edge states. For this purpose, we use a simple tight-binding model, where each atom hosts one  $p_z$-like orbital.\cite{Katsnelson} The model can be used to describe both valence- and conduction-band edges in black phosphorus. For a single layer, we can write
\begin{align}
H&=\sum_{i\neq j}t_{i,j}^{\mathbin{\|}}c^\dagger_ic_j,
\end{align} 
where $c^\dagger_i(c_i)$ creates (annihilates) an electron at site $i$, and $t_{i,j}^{\mathbin{\|}}$ denotes the hopping parameter from site $i$ to site $j$. In Fig.~4(c), we plot the resulting band structure along the high-symmetry lines connecting the $X-\Gamma-Y$ points, evaluated for the hopping parameters  obtained from first-principles calculations.\cite{Katsnelson} This parametrization includes in total ten different parameters, although qualitatively only  the two nearest-neighbor-hopping parameters $t_1^{||}=-1.486$eV and $t_2^{||}=3.729$eV, shown in Fig.~4(a), are required. 
Inspection of Fig.~4(b) indeed confirms that phosphorene exhibits $\mathcal{I}$ symmetry around the center of the unit-cell. In particular, if we label the inequivalent sites in the unit-cell as in Fig.~4(b), we find that $\hat{I}=\sigma_x\otimes\sigma_x$. The $\pm$ signs in the band structure denote the eigenvalues of $\hat{I}$. Using these eigenvalues, we can easily calculate the $\mathbb{Z}_2$ invariants $\chi_x$ and $\chi_y$ associated with edges along the $x$ and $y$ directions, respectively; we find $\chi_x=\chi_y=+1$. From this result, we can immediately infer that both bearded and armchair phosphorene  [see Fig.~4(b)], do not feature edge states. However, zigzag-terminated phosphorene cannot be tiled with an integer number of unit-cells. Therefore, we cannot simply infer the presence of edge states from $\chi_y$. Instead, Eq.~(7) states that we need to account for the ionic charge $e$ contained in the broken unit cell at the edge, see Fig.~4(b). The edge charge is given by $\sigma(k_{||})=e\gamma(k_{||})/\pi+e=e$. Hence, we expect the presence of edge states for zigzag-terminated phosphorene. These conclusions are confirmed by plotting the three different spectra, see Figs. 4(d), (e), and (f).  Here, only zigzag-terminated phosphorene exhibits one pair of in-gap edge states, which can be attributed to the edge charges (shown in red).

We can repeat this analysis for {\it bilayer} black phosphorus, for which the structure is shown in Figs.~5(a) and (e). Note that the second layer is displaced by half a lattice vector in the zigzag direction, but the stacking respects the $\mathcal{I}$ symmetry. One can also study this system using a tight-binding model
\begin{align}
H&=\sum_{i\neq j}t_{i,j}^{\mathbin{\|}}c^\dagger_ic_j+\sum_{i\neq j}t_{i,j}^\perp c^\dagger_ic_j,
\end{align} 
where we have now included interlayer hopping $t_{i,j}^\perp$. In Fig.~5(a), we have pictured the dominant interlayer hopping $t_1^\perp=0.524$eV. The resulting bulk-band structure is shown in Fig.~5(i). The two bulk $\mathbb{Z}_2$ invariants are trivial, $\chi_x=\chi_y=+1$. This implies that both bearded and armchair bilayer ribbons do not exhibit edge states. For zigzag bilayer phosphorene, we still need to account for the ionic charge in the broken unit-cell at the edge. Inspection of Fig.~5(e) reveals that this contribution is equal to $2e$. However, the surface charge is only well defined modulo $2e$, and thus we expect that all three terminations are topologically trivial.  This is confirmed by plotting the three spectra, see Figs. 5(b), (c), and (d). Both armchair and bearded bilayer phosphorene do not feature any in-gap states, whereas the zigzag terminated one exhibits two pairs. Although bearded and zigzag terminated bilayer phosphorene feature very different edge physics, they are topologically identical, owing to the  $\mathbb{Z}_2$ nature of the invariant.

It has been noticed in {\it ab initio} calculations that a potential bias $\Delta V$ applied between the two layers can drastically affect the band structure.\cite{Zunger} In particular, one can induce a Lifshitz transition if $\Delta V$ exceeds a critical value. Then, the valence and conduction band invert, and a band-crossing point emerges along the line connecting the $\Gamma-Y$ points, whereas a small gap is opened along the line connecting the $\Gamma-X$ points, see Fig.~5(j). This band inversion is accompanied by a topological phase transition, such that now $\chi_x=\chi_y=-1$. Hence, by varying this bias potential one transforms an insulator with a trivial $\mathbb{Z}_2$ invariant, as in Table~II(a), into a semimetal with a non-trivial $\mathbb{Z}_2$ invariant, as in Table~II(d). In Figs.~5(f) and (g), we plot the spectra for zigzag and bearded bilayer phosphorene. Here, one can indeed see the in-gap edge states located between the two band-crossing points [see also the close-ups in Figs.~5(k) and (l)]. These spectra are qualitatively similar to the band structure shown in Table~II(d). However, for the ribbon with armchair termination, the spectrum does not exhibit any edge states, see Figs.~5(h) and (m). This behavior is grounded on the fact that for this termination, the two band-crossing points coincide at $k_{||}=0$. This example provides a good illustration of our claim that edge states are a robust feature of Dirac semimetals, and that their existence can be attributed to the Zak phase $\gamma(k_{||})$, which changes from $\pi$ to zero as one traverses the band-crossing point. Although we have limited ourselves to single and bilayer black phosphorus, our conclusion can easily be generalized to other few-layer configurations. 
\section{Conclusion and Discussion}
In conclusion, we show that the interplay between $\mathcal{T}$- and $\mathcal{I}$-symmetry gives rise to a topological $\mathbb{Z}_2$ invariant $\chi_1$, which is directly related to the quantization of the Zak phase $\gamma(k_{||})$ in both insulators and semimetals. In particular, we find that a non-trivial Zak phase generally leads to edge states. Hereby, we have generalized the result by Ryu and Hatsugai~[\onlinecite{Hatsugai}] to systems lacking chiral symmetry. Moreover, we have extended the usual classification of 2D $\mathcal{I}$-symmetric insulators given in Ref.~[\onlinecite{Bernevig,Zaanen}].

These results are relevant for a broad range of 2D materials, including graphene, phosphorene and their multi-layer configurations. Our results explain the robust topological origin of edge states in Dirac semimetals, due to the $\pi$ Berry phase of the Dirac cone. This work, therefore, complements  earlier studies on edge states in graphene\cite{Akhmerov} based on the Dirac equation. We note that silicene,\footnote{Since the SOC in silicene is of the order of $20$K, STM experiments must be performed at $4$K to resolve the gap. At room temperature the material can be considered, for all effects, as gapless, and hence can be described by the theory presented here.} germanene, stanene, and transition-metal dichalcogenides, which also exhibit similar properties to the previous materials, are excluded from our analysis because of a significant SOC and/or the lack of $\mathcal{I}$-symmetry. 

Experimentally, the presence of edge states may be most easily detected via scanning-tunneling microscope (STM) experiments, which probe the local density of states.\cite{vandenbrink} For an insulator, the excess density of states at the surface will be quantized, whereas for semimetals it will be proportional to the  distance between the two band-crossing points in the reduced 1D BZ.\cite{Delplace} This is particularly relevant for few-layer phosphorene, where a gate voltage can induce an insulator to semimetal transition.\cite{Zunger} In the semimetallic regime, the gate voltage controls the distance in momentum space between the two band-crossing points, and as such it provides new experimental possibilities to verify our predictions. Experimentally, this insulator to semimetal transition has already been realized by depositing potassium atoms on black phosphorus.\cite{Science,nanolet} Therefore, we hope that our work will motivate future STM experiments in few-layer black phosphorus. In order to avoid contamination, one should cleave the black phosphorus and perform the STM experiments  in an ultra-high vacuum environment.\cite{Science}

We would still like to comment on the role  of disorder. It has been shown that in 1D insulators, the surface charge is immune to disorder near the edges.\cite{Loss} Moreover, the edge charges are stable against small amounts of disorder in the bulk, which preserve $\mathcal{I}$-symmetry on average.\cite{Schnyder,Diez}  

Finally, we would like to point out that the relevance of our results is not restricted to 2D materials because the Zak phase has been recently used to explain the existence of drumhead surface states in the three-dimensional materials Cu$_3$N and Ca$_3$P$_2$.\cite{Schnyder,Rappe} 
\section{ ACKNOWLEDGEMENTS}
G.v.M. and C.M.S. acknowledge financial support from NWO and the Dutch FOM association with the program ”Designing Dirac carriers in semiconductor honeycomb lattices”. C.O. acknowledges the financial support of the Future and Emerging Technologies (FET) programme within the Seventh Framework Programme for Research of the European Commission under FET-Open grant number: 618083 (CNTQC), and Deutsche Forschungsgemeinschaft under Grant No. OR 404/1-1. This work is part of the D-ITP consortium, a program of the Netherlands Organization for Scientific Research (NWO) that is funded by the Dutch Ministry of Education, Culture and Science (OCW).
\appendix
\begin{widetext}
\section{Properties of Berry connection and curvature}
In the presence of $\mathcal{T}$-symmetry, the Berry connection is even up to a total derivative. For simplicity, we consider the case where the $\mathcal{T}$-operator is represented by complex conjugation $\mathcal{K}$. Then,
\begin{align*}
A(k)&=i\langle u_k|\nabla_k|u_k\rangle=i\langle \mathcal{K}u_{-k}| e^{-i\phi(k)}\nabla_k e^{i\phi(k)} |\mathcal{K}u_{-k}\rangle=i\langle \mathcal{K}u_{-k}|\nabla_k|\mathcal{K}u_{-k}\rangle-\nabla_k\phi(k)\\
&=-i\langle u_{-k}|\nabla_k|u_{-k}\rangle-\nabla_k\phi(k)=i\langle u_{-k}|\nabla_{-k}|u_{-k}\rangle-\nabla_k\phi(k)=A(-k)-\nabla_k\phi(k).
\end{align*}
In the presence of $\mathcal{I}$-symmetry, the Berry connection is odd up to a total derivative,
\begin{align*}
A(k)&=i\langle u_k|\nabla_k|u_k\rangle=i\langle u_{-k}|e^{-i\chi(k)}\hat{I}\nabla_k\hat{I}e^{i\chi(k)}|u_{-k}\rangle=i\langle u_{-k}|\nabla_k|u_{-k}\rangle-\nabla_k\chi(k)\\
&=-i\langle u_{-k}|\nabla_{-k}|u_{-k}\rangle-\nabla_k\chi(k)=-A(-k)-\nabla_k\chi(k).
\end{align*}
Hence, from this it follows that the polarization is quantized for an $\mathcal{I}$-symmetric system, as $\int A(k)=-\int A(-k)+\int\nabla\chi(k)=-\int A(k)+2\pi j$ where the integral is over a symmetric domain. In the presence of both $\mathcal{I}$ and $\mathcal{T}$-symmetry, the integral along any closed contour is quantized, as we find that $2A(k)=\nabla\chi(k)$.\\
Finally, at the level of the Berry curvature,$\mathcal{T}$-symmetry dictates $F(k)=-F(k)$, as the curl of an even function is odd, and $\mathcal{I}$ dictates $F(k)=F(-k)$ as the curl of an odd function is even. Hence, when both $\mathcal{T}$ and $\mathcal{I}$-symmetry are present, we find $F=0$.     
\section{Relation between the Zak phase and eigenvalues of inversion}
The Zak phase is defined as $\gamma=i\int\mathrm{d}k\langle u_k|\nabla_k|u_k\rangle$, where $u_k$ is the periodic part of the full Bloch wave function $\Psi_k$. In a tight-binding model, we find $u_{k,j,\alpha}=e^{-i k( j a+r_\alpha)}\Psi_{k,j,\alpha}$, where $j$ labels the unit-cells, $\alpha$ is an orbital index, and $r_{\alpha}$ the corresponding location with respect to the center of the $j$th unit-cell. Note that the inner product is restricted to one unit cell. Furthermore, the Zak phase should be calculated for the periodic gauge, which in terms of the full wave function, translates into $\Psi_k=\Psi_{k+2\pi/a}$. Using the relation between $\Psi_k$ and $u_k$, we can then rewrite the Zak phase as
\begin{align*}
\gamma&=i\int\mathrm{d}k \langle\Psi^*_{k}|\nabla_k|\Psi_{k}\rangle+\sum_\alpha\int\mathrm{d}k |\Psi_{k,\alpha}|^2 r_\alpha 
\end{align*}
The second term on the right-hand side vanishes in the presence of $\mathcal{I}$-symmetry, since the charge distribution is symmetric around the center of the unit cell. Hence, we can write $\gamma=i\int\mathrm{d}k \langle\Psi^*_{k}|\nabla_k|\Psi_{k}\rangle$. Then, for a system without degeneracies, $\mathcal{I}$-symmetry guarantees that $|\Psi_k\rangle=e^{-i\phi(k)}\hat{I}|\Psi_{-k}\rangle$, with $\phi$ some arbitrary phase. Hence, we can rewrite the Zak phase as
\begin{align*}
\gamma&=i\int_0^\pi\mathrm{d}k\langle \Psi_k|\nabla_k|\Psi_k\rangle+i\int_{-\pi}^0\mathrm{d}k\langle \Psi_k|\nabla_k|\Psi_k\rangle=i\int_0^\pi\mathrm{d}k\langle \Psi_k|\nabla_k|\Psi_k\rangle+i\int_{-\pi}^0\mathrm{d}k\langle \Psi_{-k}|\hat{I}^\dagger e^{i\phi(k)}\nabla_ke^{-i\phi(k)}\hat{I}|\Psi_{-k}\rangle\\
&=i\int_0^\pi\mathrm{d}k\langle \Psi_k|\nabla_k|\Psi_k\rangle+i\int_{-\pi}^0\mathrm{d}k\langle \Psi_{-k}|\nabla_k|\Psi_{-k}\rangle+\int_{-\pi}^0\mathrm{d}k\nabla_k\phi(k)\\
&=i\int_0^\pi\mathrm{d}k\langle \Psi_k|\nabla_k|\Psi_k\rangle+i\int_{\pi}^0\mathrm{d}k\langle \Psi_{k}|\nabla_k|\Psi_{k}\rangle+\int_{-\pi}^0\mathrm{d}k\nabla_k\phi(k)=\phi(0)-\phi(-\pi).
\end{align*}
In the penultimate step, we used that $\nabla_k=-\nabla_{-k}$. Although the phase $\phi(k)$ is arbitrary for generic $k$, this is not true for $\mathcal{I}$-invariant momenta, as follows from the identity
\begin{align*}
|\Psi_{k_\textrm{inv}}\rangle=e^{i\phi(k_\textrm{inv})}\hat{I}|\Psi_{k_\textrm{inv}}\rangle=e^{i\phi(k_\textrm{inv})}\xi(k_\textrm{inv})|\Psi_{k_\textrm{inv}}\rangle.
\end{align*}
Hence, $\phi(k_\textrm{inv})=2\pi j+[\xi(k_\textrm{inv})-1]\pi/2$, and thus
\begin{align*}
\gamma=\phi(0)-\phi(-\pi)=2\pi (j-j')+[\xi(0)-\xi(-\pi)]\pi/2.
\end{align*}
\section{parameters for the two-band toy model}
Table~III lists the hopping parameters that have been used to obtain the spectra shown in Table~II. 
\begin{table*}[h!]
\begin{tabular}{|l||c|c|c|c|}
\hline
&(a)&(b)&(c)&(d)\\
\hline
\hline
$e_s$&-3&-5&-6&-3\\
\hline
$e_p$&1&3&7&4\\
\hline
$t_{1s}$&-2&-2&-1&3\\
\hline
$t_{2s}$&3&3&2&2\\
\hline
$t_{1p}$&1&4&2&-1\\
\hline
$t_{2p}$&-4&-3&-1&-3\\
\hline
$t_{1sp}$&-2&-2&-2&-2\\
\hline
$t_{2sp}$&4&3&1&3\\
\hline
\end{tabular}
\caption{Parameters used for the band structures from Table II.}
\end{table*}
\section{}
For the band structure shown in Fig.~2(a), we have used a tight-binding model that includes  long-range hopping. The Fourier transformed bulk Hamiltonian reads
\begin{align*}\
H(\vec{k})=h_I(\vec{k})\mathds{1}+h_y(\vec{k})\sigma_y+h_z(\vec{k})\sigma_z,
\end{align*}
with
\begin{align*}
h_{I}(\vec{k})+h_z(\vec{k})&=-0.2-0.46\cos(\vec{k}\cdot\vec{a}_1)+2.15\cos(\vec{k}\cdot\vec{a}_2),\\
h_I(\vec{k})-h_z(\vec{k})&=-0.52-0.29\cos(\vec{k}\cdot\vec{a}_1)-0.58\cos(\vec{k}\cdot\vec{a}_2)+0.6\cos(2\vec{k}\cdot\vec{a}_2)+0.3\cos(\vec{k}\cdot(\vec{a}_1+2\vec{a}_2))\\
&+0.3\cos(\vec{k}\cdot(-\vec{a}_1+2\vec{a}_2)),\\
h_y(\vec{k})&=1.81\sin(\vec{k}\cdot \vec{a}_2).
\end{align*}
\end{widetext}


\begin{thebibliography}{10}
\bibitem{HasanKane}
\textrm{M. Z. Hasan and C. L. Kane}, {Rev. Mod. Phys.} {\bf 82}, 3045 (2010).
\bibitem{QiZhang}
\textrm{ X.-L. Qi and S.-C. Zhang}, {Rev. Mod. Phys.} {\bf 83}, 1057 (2011).
\bibitem{AZ}
\textrm{A. Altland and M. R. Zirnbauer}, {Phys. Rev. B} {\bf 55}, 1142 (1997).
\bibitem{Ryu}
\textrm{A.P. Schnyder, S. Ryu, A. Furusaki, and A.W.W. Ludwig}, {Phys. Rev. B} {\bf 78}, 195125 (2008).
\bibitem{Kitaev}
\textrm{A. Kitaev}, {AIP Conf. Proc.} {\bf 1134}, 22 (2009).
\bibitem{Schnyder2}
\textrm{S. Ryu, A.P. Schnyder, A. Furusaki, and A.W.W. Ludwig}, {New J. Phys.} {\bf 12}, 065010 (2010).
\bibitem{TKNN}
\textrm{D. J. Thouless, M. Kohmoto, M. P. Nightingale, and M. den Nijs}, {Phys. Rev. Lett.} {\bf 49}, 405 (1982).
\bibitem{KaneMele}
\textrm{C. L. Kane and E. J. Mele}, {Phys. Rev. Lett.} {\bf 95}, 146802 (2005).
\bibitem{Laughlin}
\textrm{R. B. Laughlin}, {Phys. Rev. B} {\bf 23}, 5632 (1981).
\bibitem{Hatsugai}
\textrm{S. Ryu and Y. Hatsugai}, {Phys. Rev. Lett.} {\bf 89}, 077002 (2002).
\bibitem{Delplace}
\textrm{P. Delplace, D. Ullmo, and G. Montambaux}, {Phys. Rev. B} {\bf 84}, 195452 (2011).
\bibitem{Ezawa}
\textrm{M. Ezawa}, {New J. Phys.} {\bf 16}, 115004 (2014).
\bibitem{Vanderbilt}
\textrm{R. D. King-Smith and D. Vanderbilt}, {Phys. Rev. B} {\bf 47}, 1651(R) (1993).
\bibitem{Zak}
\textrm{J. Zak}, {Phys. Rev. Lett.} {\bf 62}, 2747 (1989).
\bibitem{Bernevig}
\textrm{T.L. Hughes, E. Prodan, B.A. Bernevig}, {Phys. Rev. B} {\bf 83}, 245132 (2011).
\bibitem{ortix}
\textrm{A. Lau, C. Ortix and J. van den Brink}, {Phys. Rev. Lett.} {\bf 115} 216805 (2015).
\bibitem{hatsugai3}
\textrm{T. Kariyado and Y. Hatsugai}, {Phys. Rev. B} {\bf 88} 245126 (2013).
\bibitem{Wegner}
\textrm{F. Wegner}, {Z. Phys. B} {\bf 36}, 209 (1980).
\bibitem{Li}
\textrm{ L. Li, Y. Yu, G. Jun Ye, Q. Ge, X. Ou, H. Wu, D. Feng, X. Hui
Chen, and Y. Zhang}, {Nat Nanotechnol} {\bf  9}, 372 (2014).
\bibitem{Buscema}
\textrm{M. Buscema, D. J. Groenendijk, S. I. Blanter, G. A. Steele, H. S. J. van der Zant, and A. Castellanos-Gomez}, {Nano Lett.} {\bf 14}, 3347 (2014).
\bibitem{Koenig}
\textrm{S. P. Koenig, R. A. Doganov, H. Schmidt, A. H. Castro Neto,
and B. \"Ozyilmaz}, {Appl. Phys. Lett.} {\bf 104}, 103106 (2014).
\bibitem{Liu}
\textrm{H. Liu, A. T. Neal, Z. Zhu, D. Tomanek, and P. D. Ye}, {ACS Nano} {\bf 8}, 4033 (2014).
\bibitem{Xia}
\textrm{F. Xia, H. Wang, and Y. Jia}, {Nat. Commun.} {\bf 5}, 4458 (2014).
\bibitem{Katsnelson}
\textrm{A. N. Rudenko, S. Yuan, and M. I. Katsnelson}, {Phys. Rev. B} {\bf 92}, 085419 (2015).
\bibitem{Zunger}
\textrm{Q. Liu, X. Zhang, L. B. Abdalla, A. Fazzio, and A. Zunger}, {Nano Lett.} {\bf 15}, 1222 (2015).
\bibitem{Zaanen}
\textrm{R.J. Slager, A. Mesaros, V. Juricic, and J. Zaanen}, {Nature Phys.} {\bf 9} 98 (2013).
\bibitem{Akhmerov}
\textrm{A. R. Akhmerov and C. W. J. Beenakker}, {Phys. Rev. B} {\bf 77}, 085423 (2008).
\bibitem{vandenbrink}
\textrm{C. Pauly, B. Rasche, K. Koepernik, M. Liebmann, M.
Pratzer, M. Richter, J. Kellner, M. Eschbach, B. Kaufmann,
L. Plucinski, C. M. Schneider, M. Ruck, J. van den Brink,
and M. Morgenstern}, {Nat. Phys.} {\bf 11}, 338 (2015).
\bibitem{Science}
{J. Kim, S. S. Baik, S. H. Ryu, Y. Sohn, S. Park, B.-G. Park, J.
Denlinger, Y. Yi, H. J. Choi, and K. S. Kim}, {Science} {\bf 349}, 723
(2015).
\bibitem{nanolet}
\textrm{S.S. Baik, K.S. Kim, Y. Yi, and H.J. Choi}, {Nano Lett.} {\bf 15}, 7788 (2015).
\bibitem{Loss}
\textrm{J.-H. Park, G. Yang, J. Klinovaja, P. Stano, D. Loss}, {arXiv:1604.05437}
\bibitem{Schnyder}
\textrm{Y.-H. Chan, C.-K. Chiu, M. Y. Chou, and A. P. Schnyder}, {Phys. Rev. B} {\bf 93}, 205132 (2016).
\bibitem{Diez}
\textrm{M. Diez, D.I. Pikulin, I.C. Fulga, and J. Tworzyd\l{}o}, {New. J. Phys.} {\bf 17}, 043014 (2015).
\bibitem{Rappe}
\textrm{Y. Kim, B.J. Wieder, C. L. Kane, and A.M. Rappe}, {Phys. Rev. Lett.} {\bf 115}, 036806 (2015).
\end{thebibliography}
\end{document}